\documentclass[final,3p,times,sort]{elsarticle}

\usepackage{amssymb}

\usepackage{algorithm}
\usepackage{algpseudocode}
\usepackage{amsmath}
\usepackage{float}
\usepackage{enumitem}
\usepackage{booktabs}
\usepackage{mathtools}
\usepackage{multirow}
\usepackage{multicol} 
\usepackage{dsfont}
\usepackage{subfigure}
\usepackage[table]{xcolor}
\usepackage{etoolbox}
\usepackage{pgf}
\usepackage{hhline} 
\usepackage{highlight}

\usepackage{hyperref}
\hypersetup{hidelinks,colorlinks,breaklinks=true,urlcolor=black,citecolor=black,linkcolor=black,bookmarksopen=false,pdftitle={Title},pdfauthor={Author}}
\usepackage{cleveref}

\usepackage{soul}
\soulregister\cite7
\soulregister\citealp7
\soulregister\ref7

\usepackage[flushleft]{threeparttable}
\usepackage{acronym}
\usepackage[acronym]{glossaries}
\usepackage{framed}
\usepackage{nomencl} 
\makenomenclature

\algnewcommand\algorithmicforeach{\textbf{for each:}}
\algnewcommand\ForEach{\item[ \algorithmicforeach]}

\journal{}

\begin{document}

\begin{frontmatter}

\title{A Hybrid Probabilistic Battery Health Management Approach for Robust Inspection Drone Operations}
\author[1]{Jokin Alcibar}
\ead{j.alcibar@mondragon.edu}
\cortext[cor1]{Corresponding author}
\affiliation[1]{organization={Mondragon University, Electronics, Computer Science Department},
    addressline={Loramendi, 4.}, 
    city={Mondragón},
    postcode={20500}, 
    country={Spain}}

\author[2,3]{Jose I. Aizpurua \corref{cor1}}
\ead{joxe.aizpurua@ehu.eus}

\author[4]{Ekhi Zugasti}
\ead{ekhi.zugasti@ehu.es}

\author[5]{Oier Peñagarikano}
\ead{oier@aleriontec.com}

\affiliation[2]{organization={University of the Basque Country (UPV/EHU), Computer Science and Artificial Intelligence Department},
    city={Donostia},
    postcode={20018}, 
    country={Spain}}

\affiliation[3]{organization={Ikerbasque, Basque Foundation for Science},
    city={Bilbao},
    postcode={48011}, 
    country={Spain}}

\affiliation[4]{organization={University of the Basque Country (UPV/EHU), Electronic Technology Department},
    city={Vitoria-Gasteiz},
    postcode={01006}, 
    country={Spain}}

\affiliation[5]{organization={Alerion Technologies},
    city={Donostia},
    postcode={20009}, 
    country={Spain}}

\begin{abstract}
Health monitoring of remote critical infrastructure is a complex and expensive activity due to the limited infrastructure accessibility. Inspection drones are ubiquitous assets that enhance the reliability of critical infrastructures through improved accessibility. However, due to the harsh operation environment, it is crucial to monitor their health to ensure successful inspection operations.
The battery is a key component that determines the overall reliability of the inspection drones and, with an appropriate health management approach, contributes to reliable and robust inspections. In this context, this paper presents a novel hybrid probabilistic approach for battery end-of-discharge (EOD) voltage prediction of Li-Po batteries. The hybridization is achieved in an error-correction configuration, which combines physics-based discharge and probabilistic error-correction models to quantify the aleatoric and epistemic uncertainty. The performance of the hybrid probabilistic methodology was empirically evaluated on a dataset comprising EOD voltage under varying load conditions. The dataset was obtained from real inspection drones operated on different flights, focused on offshore wind turbine inspections. The proposed approach has been tested with different probabilistic methods and demonstrates 14.8\% improved performance in probabilistic accuracy compared to the best probabilistic method. In addition, aleatoric and epistemic uncertainties provide robust estimations to enhance the diagnosis of battery health-states.
\end{abstract}




\end{frontmatter}


\section{Introduction}

Unmanned Aerial Vehicles (UAVs) have gained prominence for various civilian tasks, such as surveillance and inspection, due to their ability to hover, take off vertically, and land, making them ideal for remote and precise operations~\citep{valavanis2015}. These drones are especially useful in harsh environments, where they enhance safety and efficiency in data collection and monitoring tasks~\citep{eleftheroglou2019}. However, the robustness and reliability of UAVs are constrained by various factors such as battery life, payload capacity, and weather conditions, which can lead to mission failure or loss of the  drone~\citep{GATTI2015}.

\begin{table*}[!t]
\setlength{\tabcolsep}{3pt}
\begin{framed}
\begin{tabular}{llll}
    \multicolumn{2}{l}{\textbf{Abbreviations}} & \multicolumn{2}{l}{\textit{Nomenclatures}}\\[10px]
    ANN & \textcolor{black}{Artificial Neural Network} &$u(t)$  &Input vector of the battery model\\[2px]
    AU & \textcolor{black}{Aleatoric Uncertainty}  &$x(t)$  &State-space of the battery model\\[2px]
    BMS & \textcolor{black}{Battery Management System}  & $y(t)$  &Output vector of the battery model\\[2px]
    CNN & \textcolor{black}{Convolutional Neural Networks} &  $V(t)$  &Total battery voltage\\[2px]
    CRPS & \textcolor{black}{Continuous Ranked Probability Score}  &$\mathcal{L}_{NN}$ & Loss Function  \\[2px]
    EOD & \textcolor{black}{End-Of-Discharge} & $X_{test}$ & Test data\\[2px]
    EU & \textcolor{black}{Epistemic Uncertainty} & $h(\cdot)$ &Prediction model\\[2px]
    FC & \textcolor{black}{Fully Connected} & $\sigma^2_A$  & Variance of aleatoric uncertainty\\[2px]
    LSTM & \textcolor{black}{Long Short-Term Memory} & $\sigma^2_E$   & Variance of epistemic uncertainty \\[2px]
    MC & \textcolor{black}{Monte Carlo} & $\hat{y}$   & The mean of predicted output value\\[2px]
    MSE & \textcolor{black}{Mean Square Error} & $\rho$  &Dropout Rate   \\[2px]
    NLL & \textcolor{black}{Negative Log-Likelihood} \\[2px] 
    PDF & \textcolor{black}{Probability Density Function}  \\[2px]
    PF & \textcolor{black}{Particle Filter} \\[2px]
    PICP & \textcolor{black}{Prediction Interval Coverage Probability} \\[2px]
    PINN & \textcolor{black}{Physics Informed Neural Network} \\[2px]
    PNN & \textcolor{black}{Probabilistic Neural Network} \\[2px]
    QGB & \textcolor{black}{Quantile Gradient Boosting} \\[2px]
    QLR & \textcolor{black}{Quantile Linear Regression} \\[2px]
    QRF & \textcolor{black}{Quantile Regression Forest} \\[2px]
    RF & \textcolor{black}{Random Forest} \\[2px]
    RNN & \textcolor{black}{Recurrent Neural Network} \\[2px]
    ROM & \textcolor{black}{Reduced-Order Model} \\[2px]
    RUL & \textcolor{black}{Remaining Useful Life} \\[2px]
    SOC & \textcolor{black}{State Of Charge} \\[2px]
    TU & \textcolor{black}{Total Uncertainty} \\[2px]
    UAV & \textcolor{black}{Unmanned Aerial Vehicle} \\[2px]
    UKF & \textcolor{black}{Unscented Kalman Filter}  \\[2px]
    UQ & \textcolor{black}{Uncertainty Quantification} \\
\end{tabular}
\end{framed}
\end{table*}

Battery Management Systems (BMS) are crucial for mitigating risks through real-time diagnostics and decision-making support, ensuring mission success through fault mitigation and path replanning~\citep{sierra2019}. However, the accurate estimation of EOD voltage in BMSs is challenging due to the unavoidable presence of various sources of uncertainty such as environmental conditions or measurement errors~\citep{sierra2018}. Accordingly, it is crucial to model the uncertainty associated with battery ageing and operation to enhance the robustness of BMS predictive methodologies \citep{Zhang2022}.


\subsection{Probabilistic Battery Health State Estimation: Literature Review}

Generally, existing methods for battery health management can be classified into physics-based, data-driven, and hybrid methods~\citep{vanem2021,demirci2024}. Physics-based approaches rely on a physical model, and in most cases, battery health-state estimations are updated with filtering and tracking algorithms, e.g. Particle Filtering (PF)~\citep{Djuric2003} or Unscented Kalman Filter (UKF)~\citep{bai2023}. In contrast, data-driven methods make use of battery degradation data to infer the capacity fade model. Finally, hybrid approaches combine the advantages of physics-based and data-driven approaches \citep{guo2015}.

\subsubsection*{\textbf{Physics-based Approaches}}

Under physics-based degradation methods, \cite{daigle2013}  presented a battery prognostics model based on electrochemistry equations and updated through UKF. As a result, the EOD predictions demonstrated high accuracy with associated uncertainty. Nevertheless, this method requires the estimation of a high number of parameters, making the optimal search more complex. \cite{Pola2015} introduced an empirical state-space model for prognostics of state of charge (SOC) and EOD employing PF. The model successfully predicted the expected discharge time and provides confidence intervals. However, its parameterization for the open circuit voltage curve is not adaptive and therefore invalid for batteries of more than one cell. 

In this direction, \cite{sierra2019} estimate the SOC and EOD time of Li-Po batteries using a model-based prognostics architecture. The proposed framework incorporates a simplified battery model using artificial evolution concepts for battery parameter estimation. Additionally, it employs a feedback correction loop adjusting the variance of the process noise to mitigate bias in state estimation. Despite their good performance, physics-based models are typically computationally expensive due to the complex differential equations involved in their formulation and thus are not well-suited for online applications. 

\subsubsection*{\textbf{Data-driven Approaches}}

Several data-driven methods have been proposed to overcome the limitations of physics-based models~\citep{ng2020}. Namely, various machine learning techniques, such as Artificial Neural Networks (ANNs), Gaussian Processes, Random Forests, and Support-Vector Machines, have been successfully applied to address EOD prediction and degradation inference~\citep{Ochella2022}. Focusing on deep learning methods for state of health, Remaining Useful Life (RUL) and EOD estimation, Convolutional Neural Networks (CNNs)~\citep{Couture2022,mitici2023,Hsu2022}, Recurrent Neural Networks (RNNs)~\citep{ZHAO2023,Guo2024,meng2023} and the combination of both methods~\citep{Mazzi2024,tang2022} have shown promising results. 

CNN and RNN models are effective at extracting temporal and sequential features. The measured data can be directly used to map the nonlinear relationships between the prediction tasks. However, the main limitations of purely data-driven models are their complexity, and limited capacity to generate robust predictions for unseen data and model uncertainty \citep{ariaschao2022,Aizpurua_23}.

In order to address the uncertainty of deep ANNs, more sophisticated methods have been developed as \cite{biggio2023}, where an encoder–decoder architecture performs the EOD prediction incorporating the quantification of model uncertainty through Monte Carlo (MC) dropout. In this respect, Bayesian methods~\citep{mishra2018,Zhang2022} or Probabilistic Neural Networks (PNN)~\citep{che2024}, are suitable to infer uncertainty. Despite such models are able to make accurate predictions and quantify model uncertainty, data-driven techniques are often considered black-box methods. 

\subsubsection*{\textbf{Hybrid Approaches}}

Hybrid methods integrate physics-of-failure and data-driven models to provide better interpretability of the predictions~\citep{FINK2020}. \cite{yeregui2023} integrated a Reduced-Order Model (ROM) with a Long Short-Term Memory (LSTM) network to estimate the SOC of Lithium-ion (Li-ion) cells. However, the uncertainty has not been quantified. \cite{Fernandez2023} presents a physics-guided Bayesian neural network applied to lateral-load tests. This hybrid approach uses Approximate Bayesian Computation to train without a traditional loss function, improving prediction accuracy and uncertainty quantification (UQ). 

Another alternative that has emerged recently is based on Physics Informed Neural Networks (PINNs) concepts, where physical equations are incorporated in the machine learning loss function aiming to perform accurate estimations. PINNs have been effectively utilized to achieve significant outcomes in various engineering domains such as computational mechanics \citep{LI2021}, computational fluid dynamics \citep{raissi2020}, material design \citep{ZHANG2021} or mechanical fault diagnosis \citep{NI2023,Feng2023_2}.\cite{sun2023} exhibit a LSTM model informed by physical information for battery health prognostics. However, uncertainty quantification has not been addressed. \cite{nascimento2023} present a hybrid physics‑informed approach that simulates dynamical responses of Li-ion batteries, implementing principle‑based equations through RNN. In this case, ROMs describe the voltage discharge and model-uncertainty is captured through multi‑layer perceptrons.


\subsection{Gap Identification \& Contribution}

Table \ref{table:sota} synthesizes recent works on battery health-state estimation and highlights the key concepts developed in this research study.

\begin{table}[!ht]
    \centering
    \caption{Summary of recent battery health-state estimation methods.}
    \setlength{\tabcolsep}{5pt}
        \begin{tabular}{c c c c c }
             \toprule
             \textbf{Ref.} & \textbf{Dataset} & \textbf{Model} & \textbf{Output}& \textbf{UQ} \\
           \midrule
            {\citep{sierra2019}}  & Operational & Physics based  & {\scriptsize *}EOD$_{t}$,SOC & Yes\\
            {\citep{sierra2018}}  & Operational & Physics based  & {\scriptsize *}EOD$_{t}$,SOC & None\\
            {\citep{bai2023}}     & Laboratory  & Physics based  & RUL      & Yes \\
            {\citep{Hong2020}}    & Laboratory  & Data-Driven    & RUL      & Yes \\
            {\citep{Hsu2022}}     & Laboratory  & Data-Driven    & {\scriptsize **}EOD$_{v}$ EOL   & None \\
            {\citep{ZHAO2023}}    & Laboratory  & Data-Driven    & SOH      & None \\
            {\citep{HAN2022}}     & Laboratory   & Data-Driven   & SOH, RUL & None \\
            {\citep{meng2023}}    & Laboratory   & Data-Driven   & SOH      & None \\
            {\citep{biggio2023}} & Laboratory   & Data-Driven    & {\scriptsize **}EOD$_{v}$     & Partial \\
            {\citep{Zhang2022}} & Laboratory   & Data-Driven  & RUL   & Yes \\
            {\citep{yeregui2023}} & Laboratory & Hybrid       & SOC   & None \\
            {\citep{sun2023}} & Laboratory     & Hybrid       & SOH   & None\\
            {\citep{nascimento2023}} & Laboratory  & Hybrid  & SOC & Yes \\
            Ours & Operational & Hybrid   & {\scriptsize **}EOD$_{v}$     & Yes\\
            \bottomrule
            \multicolumn{5}{l}{\scriptsize{Legend: * EOD$_{t}$: EOD time. ** EOD$_{v}$: EOD voltage.}} \\
        \end{tabular}
        \label{table:sota}
\end{table}	

It can be observed from Table \ref{table:sota} that there are few studies that focusing on analyzing field drone battery data operated under varying operation conditions~\citep{sierra2019,sierra2018}. The majority of studies employ battery data derived from discharge processes taking place in a controlled laboratory environment~\citep{bai2023,Hong2020,Hsu2022,ZHAO2023,HAN2022,meng2023,biggio2023, Zhang2022,yeregui2023,sun2023,nascimento2023}. Furthermore, many analytic methods do not quantify uncertainty~\citep{sierra2018,Hsu2022,ZHAO2023,HAN2022,meng2023,yeregui2023,sun2023} or distinguish between data uncertainty and model uncertainty~\citep{biggio2023}, except Bayesian inference based approaches~\citep{sierra2019,bai2023,Hong2020,Zhang2022,nascimento2023,Fernandez2023}. In summary, this study covers a gap in the literature by developing a hybrid probabilistic model that quantifies uncertainty utilizing real-world operational battery data.

Accordingly, the main contribution of this paper is the development of a novel and robust hybrid probabilistic approach for EOD voltage predictions in Li-Po batteries. The novelty of the proposed approach lies in the combination of MC dropout to model epistemic uncertainty and Negative Log-Likelihood (NLL) loss function to capture aleatoric uncertainty through a hybrid residual architecture. The residual hybridization is achieved in an error-correction configuration, which combines physics-based discharge and probabilistic error-correction models. In particular, the physics-based model, rooted in electrochemistry principles, facilitates the extraction of informative features of the battery behavior without requiring a detailed model of the physics, thus avoiding high computational costs. In addition, the use of physics-based model reduces the data requirements for probabilistic data-driven models and deals with more complex scenarios, increasing efficiency. 

Subsequently, the probabilistic error-correction is developed through the explicit integration of uncertainty via CNN and MC dropout. A benchmarking comparison in terms of accuracy and uncertainty has been performed among different probabilistic models including QLR, QRF, and QGB so as to evaluate their predictive accuracy and robustness.

Detailed and explicit uncertainty modelling enables an improved understanding of the system uncertainty, which is shown to impact battery health-state estimation activities. Namely, the proposed uncertainty-aware EOD voltage prediction approach enables capturing inherent data variability and model uncertainty, and therefore, contributes to building robust EOD voltage predictions. This is demonstrated through uncertainty quantification  in different batteries, followed by a post-processing stage to diagnose the battery health-state.


The proposed methodology effectively addresses diverse battery parameter errors, such as calibration errors of physics-based battery model or variations due to electrolyte composition and electrode materials. This approach is particularly useful for drone fleet operators, enabling accurate prediction of EOD voltages across diverse drone and battery configurations. The benefits and impact of the proposed approach are demonstrated with an industrial dataset obtained from several real inspection drone flights using various manoeuvres under different loading conditions.

\subsection{Outline}

The remainder of this paper is organized as follows. Section~\ref{sec:Background} introduces a theoretical background in uncertainty quantification concepts used throughout the paper.  Section~\ref{sec:PrognosticsArchitecture} presents the proposed novel hybrid probabilistic EOD voltage developed using CNN MC dropout. Section~\ref{sec:CaseStudy} describes the case study that has been used to test the proposed approach. Section~\ref{sec:Results} presents the performance of the architecture and evaluates the results. Section~\ref{sec:Discussion} discusses the obtained results, and finally, the main conclusions are drawn in Section~\ref{sec:Conclusion}.

\section{Background} 
\label{sec:Background}

Battery EOD voltage predictions can be characterized through the accuracy and uncertainty. Uncertainty represents the prediction reliability~\citep{Heo2018}, \textit{i.e.} higher accuracy and higher uncertainty models may be less reliable than a model with lower accuracy and lower uncertainty~\citep{rathnakumar2023}. 

In order to build robust and reliable predictive models, understanding and modelling different sources of uncertainty is crucial. Within the scope of uncertainty quantification, aleatoric and epistemic uncertainty are generally defined~\citep{der2009}. In order to capture the probabilistic nature of aleatoric and epistemic uncertainties the inference of probability density functions (PDFs) over deterministic point estimates is essential~\citep{hullermeier2021}.

\subsection{Aleatoric uncertainty} 
\label{ss:Aleatoric}

Aleatoric uncertainty is a form of statistical uncertainty that is irreducible. It refers to the inherent randomness in natural processes, such as sensor noise~\citep{kendall2017}. The numerical characterization of aleatoric uncertainty involves the acquisition of knowledge about the conditional distribution of the target variable given the values of input variables~\citep{Tagasovska2019}. Aleatoric uncertainty can be further classified into homoscedastic and heteroscedastic uncertainty~\citep{kendall2017}. Homoscedastic uncertainty occurs when the data-variance have uniform distribution across the entire range~\citep{Le2005}. Heteroscedastic uncertainty involves varying levels of variance in the data. In this case, the variability of residuals varies across different conditions~\citep{Le2005}. Heteroscedasticity can be expressed as:

\begin{equation}
\label{eq:heteroscedasticity}
    Var(\epsilon|X) = \frac{1}{T} \sum_{t=1}^{T} \hat{\sigma}_t^2(X)
\end{equation}

\noindent where $\sigma^2 (X)$ indicates that the variance of residuals is dependent on $X$.

Understanding and appropriately addressing heteroscedasticity is crucial to ensure valid and reliable results, especially when working with real-world data that often exhibit varying levels of uncertainty across different ranges of predictors~\citep{gal2016}.

\subsection{Epistemic uncertainty}
\label{ss:EpistemicUncertainty}

Epistemic uncertainty denotes to the uncertainty associated with the lack of specific knowledge, which is reflected in the model-uncertainty. This uncertainty can be reduced with additional knowledge acquired through more information, \textit{i.e.} more data or improved models~\citep{Tagasovska2019}. 

\noindent Gaining knowledge about unexplored regions of the input space is an essential aspect to reduce epistemic uncertainty. Accordingly, the majority of methods rely on assessing the differences between various models trained on identical data~\citep{Lakshminarayanan2017}. This strategy enables capturing the uncertainty associated with the model architecture including hyperparameters~\citep{hullermeier2021}. Epistemic uncertainty is represented as~\citep{kendall2017}:


\begin{equation}
    Var(y) = \frac{1}{T} \sum_{t=1}^{T}f(x,\theta_t)^2 -  \left(  \frac{1}{T} \sum_{t=1}^{T}f(x,\theta_t) \right)^2
\end{equation}
 
\noindent where $f(x,\theta_t)$ denotes the output of the model for the $t$-th sample, $x$ signifies the sample in question and $\theta$ represents the parameters of the model.


\subsection{Total predictive uncertainty}
\label{ss:TotalUncertainty}


The combination of epistemic uncertainty (EU) and aleatoric uncertainty (AU), into a unique total uncertainty (TU) can be represented as follows~\citep{abdar2021}: 

\begin{equation}
\label{eq:predictiveUncertaintyRoot}
    TU = \sqrt{EU^2 + AU^2}
\end{equation}

Taking the square root of the sum of squared aleatoric and epistemic uncertainties helps to ensure that the total uncertainty is in the same units as the individual uncertainties and follows the principles of Euclidean geometry use to calculate the magnitude for independent components.

\section{Uncertainty-Aware Battery Health Management Approach}
\label{sec:PrognosticsArchitecture}
\begin{figure*}[!htb]
    \centering
    \includegraphics[width=1\textwidth]{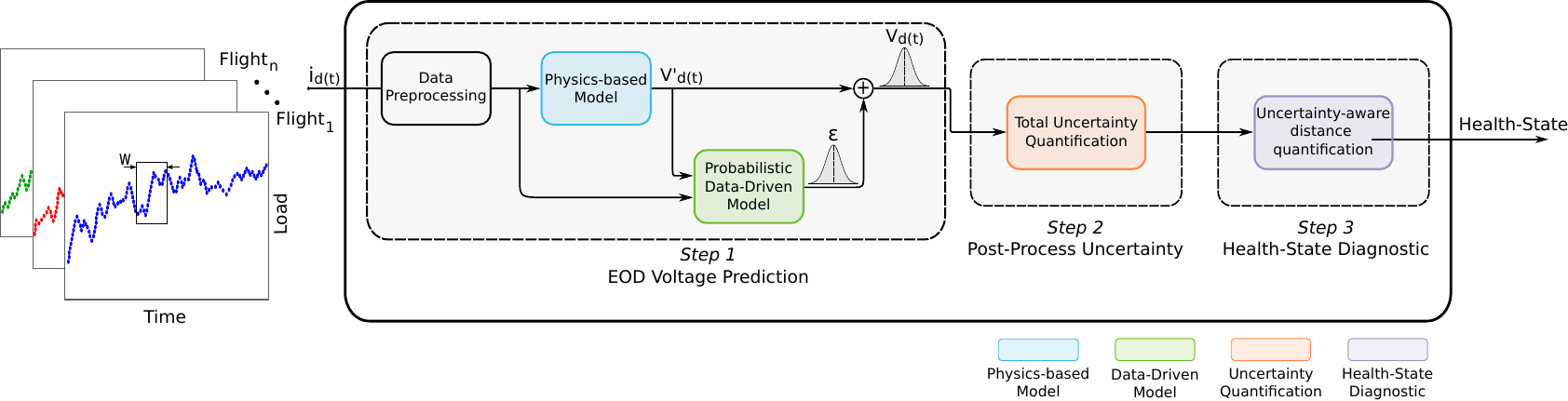}
    \caption{Uncertainty-aware battery health management approach.}
    \label{fig:prog_architecture}
\end{figure*}

Figure~\ref{fig:prog_architecture} shows the proposed uncertainty-aware battery health management approach constituted of EOD voltage prediction followed by post-process UQ  and health-state diagnostic stages.

\subsection{Step 1: EOD Voltage Prediction}
\label{ss:EODV_prognostics}


This section explains Step 1 of Figure~\ref{fig:prog_architecture}. The EOD voltage prediction approach is comprised of a physics-based and a data-driven model connected in a error-correction configuration. A preprocessing stage is performed before the main steps. During this stage, monitored data was cleaned by removing variables without data, doubled records, and incorrect sensor readings. Missing records were imputed with average values to ensure data completeness and consistency.

The physics-based approach defines the physics-of-discharge of the battery. By taking the load at instant $t$, denoted as $i_d(t)$, the model computes an estimated terminal voltage $v'_d(t)$. This estimation is used as input to the probabilistic data-driven model in the subsequent error prediction stage. Finally, the last stage focuses on the dynamic error-correction to estimate the discharge voltage $v_d(t)$.

\subsubsection{Physics-based discharge model}
\label{ss:PhysicsBased}

The electrochemistry-based battery discharge model is inspired from \cite{daigle2013} and implemented through the Prognostics Model Library~\citep{2023_nasa_progpy} from NASA Ames Research Center. This model operates using a set of ordinary differential equations to determine the total battery voltage, $V(t)$, which is defined as the difference in potential between the positive and negative current collectors, accounting for the resistance losses come across these current collectors. The state-space of the battery model, $x(t)$, input vector, $u(t)$, and output vector, $y(t)$ are defined by Eqs. (\ref{eq:state_vector}) - (\ref{eq:output_vector}) respectively~\citep{salinas-camus2023}:

\begin{gather} 
x(t) = [q_{s,p}, q_{b,p}, q_{s,n}, q_{b,n}, V', V'_{n,p}, V'_{n,n}]^T \label{eq:state_vector} \\ 
u(t) = [i_{app}] \label{eq:input_vector}\\ 
y(t) = [V] \label{eq:output_vector}
\end{gather}

The state-space includes $q$ and $V'$, corresponding to the charge on the electrodes and the voltage differentials, respectively. The notations $p$ and $n$ indicate the positive and negative electrodes, which are divided into two volumes representing the surface layer ($s$) and  the bulk layer ($b$). The input vector, indicates the electric current applied to the battery and the output vector denotes the discharge voltage of the battery.

\subsubsection{Probabilistic Data-driven model}
\label{ss:DataDriven}
For the error-correction stage, a probabilistic CNN has been developed due to its feature extraction and uncertainty estimation capability. Identifying and quantifying uncertainty in drone operations is crucial. Aleatoric uncertainty increases due to the inherent randomness within the system, such as wind variability, sensor errors, and diverse operational conditions. Conversely, epistemic uncertainty arises from a lack of knowledge, including modeling errors. Therefore, distinguishing between these types of uncertainties is essential for effective decision-making, as it enables operators to mitigate risks associated with uncertain information.\\

\noindent \textbf{CNN MC dropout}
\label{ss:CNN_Architecture}

\begin{figure}[ht]
    \centering
    \includegraphics[width=0.99\textwidth]{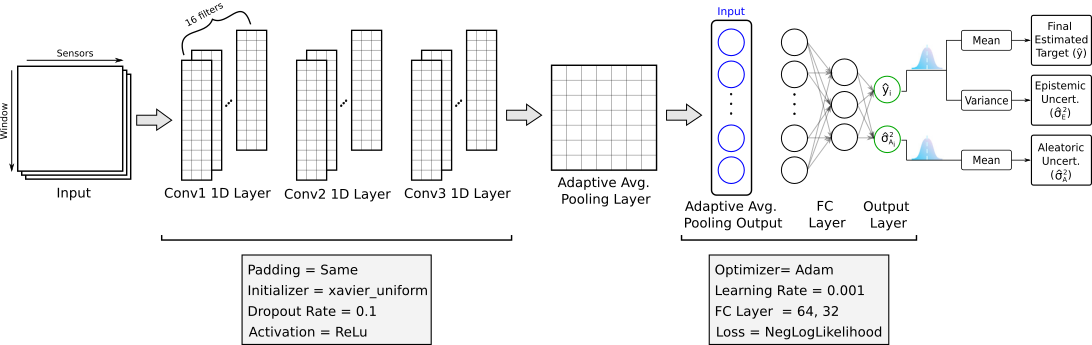}
    \caption{Structure of the proposed uncertainty-aware CNN architecture}.
    \label{fig:cnn_architecture}
\end{figure}

CNNs are deep learning models specifically designed for processing data structured in a tensor format. They demonstrate proficiency in handling high-dimensional data and can automatically capture underlying features for making accurate predictions~\citep{xu2022}. The use of CNN with MC dropout demonstrates that integrating dropout during both training and testing phases, significantly improves the robustness of CNNs providing a more comprehensive probabilistic interpretation of model predictions \citep{mitici2023,choubineh2023}. The structure of the proposed CNN based error-correction model is shown in Figure~\ref{fig:cnn_architecture}, and a comprehensive description of each layer is outlined below.

\begin{itemize}[leftmargin=15pt]
    \item Input data: the input data for the CNN is structured in a tensor format. The rows represent data samples acquired through windowing, and columns that correspond to sensors deployed on the inspection drone, such as the current load and the estimated voltage derived from a physics-based model.

    \item Convolutional layers (Conv1 to Conv3 in Figure~\ref{fig:cnn_architecture}): these layers are composed of Conv1D layers, which consist of an array of convolutional filters used for extracting features from the input data, especially in the analysis of sequential data. These filters are utilized on the input sequence to identify unique patterns associated to the prediction variable. 
    In each convolutional layer, a total of 16 filters were applied to generate the feature maps. This number was selected after testing with 8, 16, and 32 filters, considering the performance and model complexity.

    

    \item Adaptive Average Pooling layer: in this layer, for each feature map, the average of all the activations is computed, resulting in a single value per feature map. This reduces the spatial dimensions of the feature maps to a scalar value, which is then used as a summary of that feature map. 

    \item Fully Connected (FC) layers: these layers serve the purpose of transforming the extracted features into predictions through the utilization of a linear activation function. Two FC layers have been used. This enables learning complex relationships and patterns from the learned features, enabling accurate predictions based on the input data.

    \item Output layer: this layer is responsible for producing the final results given the inputs and the learned weights from the previous layers. The output layer consists of two neurons representing the mean and variance, in order to quantify the expected value and its associated uncertainty. To ensure a positive variance, the neuron is activated using an exponential function.
\end{itemize}

\begin{table}[ht]
    \centering
    \caption{Architectural and Optimization Hyperparameters of CNN.}
    \begin{tabular}{p{2.3cm} p{8cm} p{4.2cm}}
        \toprule
        \textbf{Category} & \textbf{Hyperparameter} & \textbf{Value} \\
        \midrule
        \multirow{9}{*}{Architecture} & Window-size & 10 \\
        & Convolutional layers & 3 \\
        & Number of filters  & 16 \\
        & Kernel size & 3 \\
        & Padding & Same  \\
        & Initializer & Xavier Uniform \\
        & Activation & Rectified Linear Unit \\
        & Number of fully connected layers & 2 \\
        & Number of nodes in the fully connected layers & 64, 32 \\
        \midrule
        \multirow{5}{*}{Optimization} & Loss Function & NLL \\
        & Optimizer & Adam \\
        & Learning Rate & 0.001 \\
        & Number of epochs & 130 \\
        & Monte Carlo dropout rate & 0.1 \\
        & Number of Monte Carlo inferences  & 100 \\
        \bottomrule
    \end{tabular}
    \label{table:cnn_hyperparameters}
\end{table}

\begin{figure}[ht]
    \centering
    \includegraphics[width=0.45\textwidth]{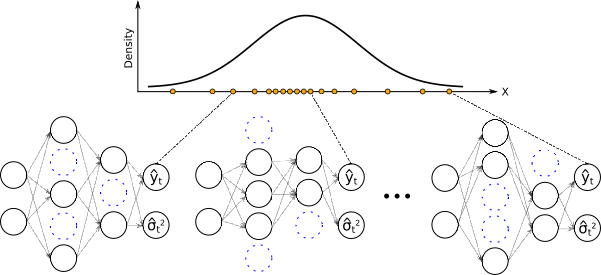}
    \caption{Use of dropout at test stage with MC sampling.}
    \label{fig:mc_dropout}
\end{figure}

The dropout layer prevents overfitting during the training phase by randomly switching-off neurons. At testing time, MC dropout technique is implemented to provide a scalable way to learn a predictive distribution~\citep{gal2016}. MC dropout works as shown Figure~\ref{fig:mc_dropout} in the training stage, where each dropout configuration corresponds to a different sample from the approximate parametric posterior distribution~\citep{gal2016}. This distribution enables the estimation of epistemic uncertainty computing the variance ($\hat{\sigma}^{2}_E$ in Figure~\ref{fig:cnn_architecture}) and contributes to enhance model robustness and accuracy.

Finally, in order to capture the aleatoric uncertainty, and assuming that the uncertainty can be modelled with an equivalent Gaussian distribution, $N(\mu,\sigma)$, the CNN architecture has been modified adding an additional output which models variance, $\sigma^2$, as shown in Figure~\ref{fig:cnn_architecture}. By modifying the architecture and employing a loss function based on NLL, the uncertainty can be learned as a function of the data~\citep{kendall2017}:

\begin{equation}
\label{eq:NegLogLiklossfuntion}
\mathcal{L}_{NN}(y_{true}|\mu,\sigma^2) = \frac{1}{N} \sum_{i=1}^{N} \frac{(y_i - \mu(x_i))^2 }{2\sigma (x_i)^2} + \frac{1}{2}log\sigma(x_i)^2 
\end{equation}

\noindent where $y_i$ is the true value for the $i$-th element in the batch, $\mu(x_i)$ is the predicted mean value for the $i$-th element and $\sigma(x_i)^2$ represents the variance of the distribution.

When the CNN accurately estimates the true value of the target variable, the prediction variance is low. This helps to minimize the mean square error (MSE) loss term in (\ref{eq:NegLogLiklossfuntion}). If there is a low signal-to-noise ratio, the CNN will estimate a higher variance, integrating the inherent uncertainty in the prediction due to the noisy data. That is, the variance (i) controls the level of uncertainty on the CNN model predictions and (ii) balances the desire to capture the inherent variability in the data with the need to avoid overly uncertain predictions.




This approach enhances the ability of the CNN model to provide accurate predictions and capture uncertainty inherent in the data, making it a valuable tool for decision-making in an inspection drone.\\

\subsection{Step 2: Post-Process Uncertainty}
\label{ss:Post-procesingUncertainty}


This section explains Step 2 of Figure~\ref{fig:prog_architecture}. The post-processing stage focuses on the inference of total uncertainty so as to improve the decision-making under uncertainty. 


\subsubsection{Total Uncertainty Quantification}
\label{sss:TUQ}

Unified modeling of aleatoric and epistemic uncertainty is crucial for a comprehensive understanding and representation of total uncertainty in a given system, process, or model~\citep{kendall2017}. To achieve this objective, the Algorithm~\ref{alg:total_uncertainty} defines the overall process of total uncertainty quantification.
            
    \begin{algorithm}[ht]
        \caption{Total Uncertainty Inference}
        \label{alg:total_uncertainty}
        \begin{algorithmic}[1]
        \State \textbf{Input:} data $X_{test}$, prediction model $h(\cdot)$, number of iterations $N$
        \State \textbf{Output:} total uncertainty $\sigma_{TU}$ 
        \For{\textbf{each} i $\in$ $N$}{}{}
            \State $[y_i,\sigma^2_{A_i}] = h(X_{test_i})$  \Comment{cf.Figure~\ref{fig:cnn_architecture}}
            \State $\vec{y}[i] = y_i$ \Comment{Store mean results}
            \State $\vec{\sigma_A}^2[i] =\sigma^2_{A_i}$ \Comment{Store variance results}
        \EndFor
        \State $\hat{y} $ = $mean(\vec{y})$ \Comment{Infer mean pred. values}
        \State $\sigma^2_E$ = $var(\vec{y})$ \Comment{Infer Ep. Uncertainty}
        \State $\sigma^2_A$ = $mean(\vec{\sigma^2_A})$ \Comment{Infer Al. Uncertainty}
        \State $\sigma_{TU}$ = $\sqrt{\sigma_E^2 + \sigma_A^2}$ 
        \State \textbf{return} $\sigma_{TU}$ 
        \end{algorithmic}
    \end{algorithm}


Summing the squares of the uncertainties combines the individual uncertainties in a way that gives greater weight to larger uncertainties. This reflects the idea that larger uncertainties contribute more significantly to the overall uncertainty in the process or measurement being considered. Finally, taking the square root of the sum brings the combined squared uncertainty back to the original scale, providing a measure of total uncertainty in the same units as the original uncertainties.

\subsection{Step 3: Battery Health-State Diagnostics}
\label{ss:BatteryHealth-StateDiagnostics}

This section explains Step 3 of Figure~\ref{fig:prog_architecture}. Drone health-management decisions are adopted based on available evidence, observations, and data. In this context, UQ holds a significant role because it captures uncertainty associated with modelling and data, and it enables decision-making under uncertainty.

In the present methodology, uncertainty analysis is employed to assess the health-state of the battery. The prediction interval for computing battery health-state can be defined based on the total uncertainty, which is trained using good health batteries. Afterwards, when new flight data for the inspection drone is obtained, (i) EOD voltage predictions are performed considering 
the uncertainty, and (ii) whole predictive information is assessed to evaluate if the observed voltage is within the predicted range. 

    \begin{figure}[!ht]
        \centering
        \includegraphics[width=0.5\textwidth]{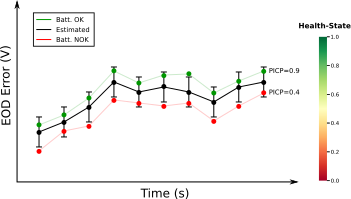}
        \caption{Health-state inference example based on prediction interval quantification.}
        \label{fig:health_state}
    \end{figure}
    

The prediction interval coverage probability metric (PICP) [cf. Eq. (\ref{eq:picp})] is used to quantify the proportion of true observations that fall within the estimated prediction interval defined by the total uncertainty. Figure~\ref{fig:health_state} shows an example of PICP estimation and inference of the battery health-state. Specifically, the green line represents instances where a significant majority of the true battery health-state values reside within the confidence intervals, which indicates that the battery under study is in a good health-state (Batt. OK). Conversely, the red line represents instances where most of the true values lie outside the confidence intervals, which indicates that the battery under study is not in a good health-state (Batt. NOK).

Finally, based on the PICP results, battery health-state diagnostic has been computed for the performed flights. This ratio is crucial in quantifying the predictive capabilities and highly significant indicator of the battery health-state diagnostic system. This diagnostic system is designed to categorize the performance of the battery health-state for each flight based on collected data. 

\section{Case Study}
\label{sec:CaseStudy}

\begin{figure}[!ht]
	\centering
	\includegraphics[height=5.3cm ,width=7cm]{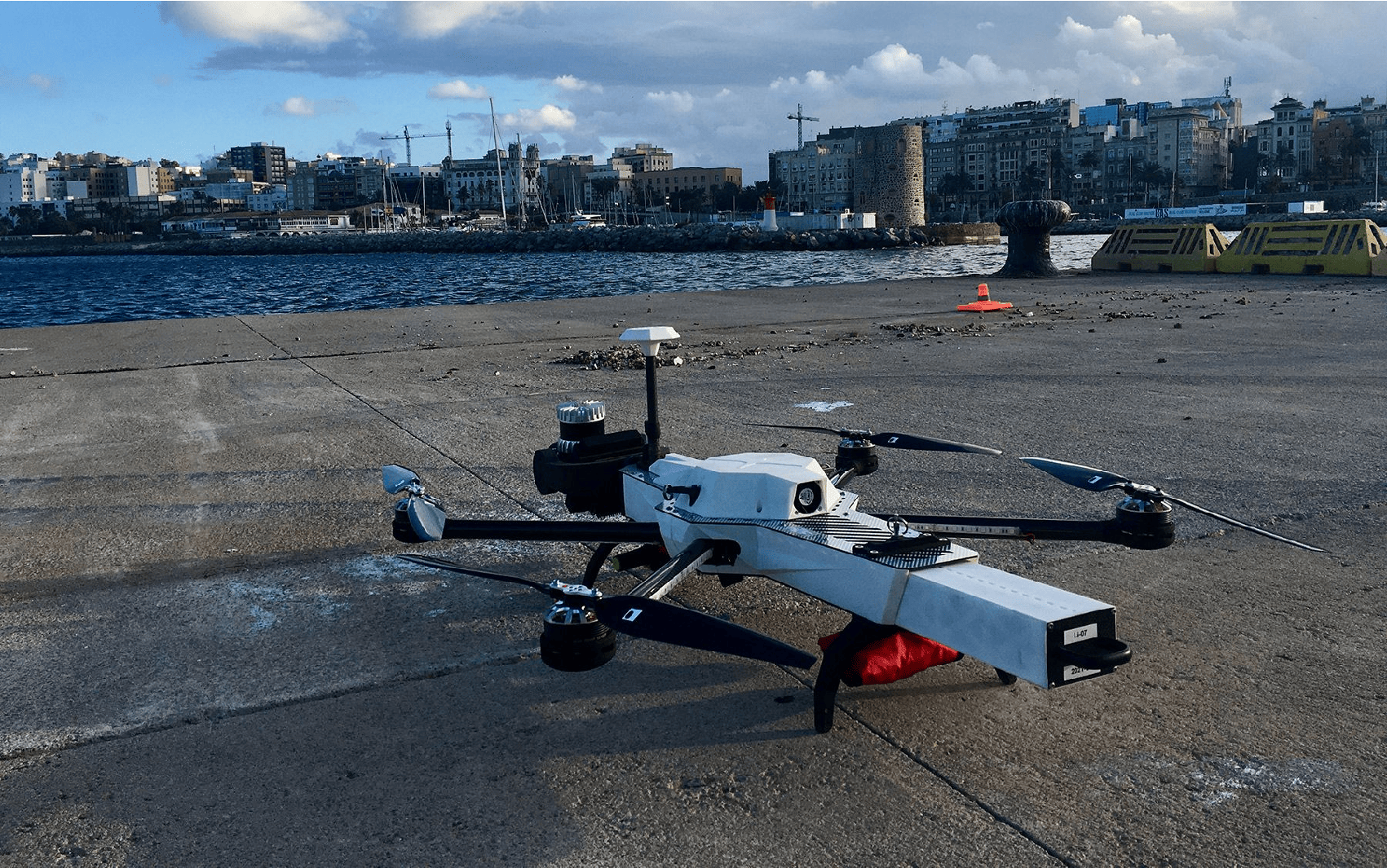}
	\caption{Offshore inspection drone.}
	\label{fig:AlerionDrone}
\end{figure}

The proposed approach is tested and validated on different flights performed by different batteries in inspection drones as show in Figure~\ref{fig:AlerionDrone}.

These drones are used for the inspection of defects in offshore wind turbines, e.g. cracks, which are processed through on-boarding processing software. These drones are all equipped with Li-Po batteries, and the focus of this research is on a single-cell of 6S, 30000 mAh, Li-Po battery.

\subsection{Description of the dataset}


For each flight, various parameters are recorded from the drone, including load and discharge voltage. Current is measured using an ACS758 Hall effect sensor, which is installed on the main positive wire to minimize the risk of short circuits on the sensor board. Voltage is measured using a voltage divider circuit equipped with a filter. This setup reduces the likelihood of a false return to launch trigger, which can occur in very windy conditions due to abrupt increases in motor speed needed to maintain level flight. There are other external variables that also affect the discharge process, such as ambient temperature or pressure. However, the focus of this paper is on the use of the loading variable to get an accurate estimate of the EOD voltage.

\begin{figure}[!ht]
	\centering
	\includegraphics[width=0.55\columnwidth]{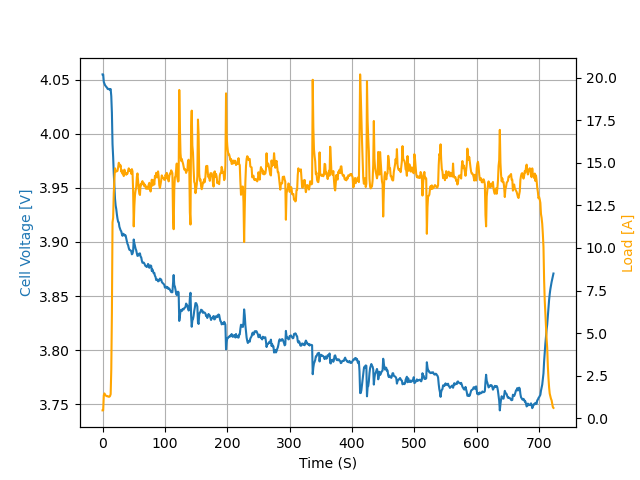}
	\caption{Available datasets for the  flight \#73.}
	\label{fig:FlightExample}
\end{figure}

\noindent Figure~\ref{fig:FlightExample} illustrates an example of the variables employed as data for each flight in the study. This reflects the difference with ideal laboratory conditions (cf. Figure~\ref{table:sota}). The overall dataset consists of 26,156 data samples obtained from 33 distinct flights. The length of each flight determines the proportion of samples that are used for validation. Accordingly, Table~\ref{table:Performance} displays the percentage of flight samples with respect to the overall dataset, which is used to validate the results for each flight. To analyze the large amount of data collected by these drones, the research uses a laptop equipped with an Intel® Core™ i5-10210U CPU (Central Processing Unit) at 1.60 GHz (Gigahertz) and 16 GB (Gigabytes) of RAM (Random Access Memory), running PyTorch software for deep learning tasks.

Table~\ref{table:dataset} provides a detailed distribution of how flight data has been divided between training and testing sets across three different years. In addition, it displays the total count of flights recorded in each year.


\begin{table}[!ht]
    \centering
    \caption{Train-test split on the flights within the dataset.}
        \begin{tabular}{c c c c c }
             \toprule
            & \textbf{2021} & \textbf{2022} & \textbf{2023}& \textbf{Total} \\
           \midrule
              \textbf{Train} & 18 & 3 & 3 & \textbf{25 (75.75\%)}   \\
              \textbf{Test} & 7 & 1 & 1 & \textbf{8 (24.25\%)}   \\
            \midrule
            \textbf{Total} & 25 & 4 & 4 & \textbf{33 (100\%)} \\
            \bottomrule
        \end{tabular}
        \label{table:dataset}
\end{table}	

The hyperparameters defined for the CNN model (Table~\ref{table:cnn_hyperparameters}) have been adjusted using the training set. Best results were obtained with the Adam optimizer, learning rate 0.001, 16 filters and 64 and 32 neurons in the two FC layers.

\subsection{Benchmarks}

In order to evaluate the predictive accuracy of the proposed CNN model, it has been compared with other probabilistic models. Namely, QLR, QRF, and QGB have been selected as these probabilistic models exhibit high performance in analogous scenarios, demonstrating robustness and accurate predictions across varied applications~\citep{song2022,Bracale2023,JIANG2023}. As with the CNN model, the hyperparameter tuning process has been done through a grid search using the training set.\\

\noindent \textbf{Quantile Linear Regression}
\label{ss:QLR}

QLR estimates conditional quantiles of a response variable, providing a full conditional distribution when traditional regression assumptions fail, and exploring relationships with predictors at different quantile levels. QLR minimizes the pinball loss function to predict observed quantiles~\citep{chung2021}. This technique provides a robust approach to modeling the conditional distribution of the response variable, making it less sensitive to outliers and skewness in the data~\citep{torossian2020}.\\

\noindent \textbf{Quantile Regression Forest}
\label{ss:QRF}

QRF combines quantile regression robustness with Random Forest (RF) principles to estimate response variable quantiles~\citep{meinshausen2006}. Unlike traditional quantile regression, which uses the pinball loss function, QRF employs an ensemble of decision trees. These trees are constructed using random selection process for node and split point determination~\citep{antoniadis2021}. For input $X = x$, QRF assigns weights $\omega_i(x)$ for each data point to compute the conditional distribution, analogous to RFs conditional mean estimation~\citep{breiman2001}. Thus, QRF provides comprehensive uncertainty quantification by inferring the weighted distribution of responses~\citep{meinshausen2006}. The optimization of hyperparameters was carried out using Bayesian optimization through \texttt{scikit-optimize}. The fine-tuning of parameters includes the \texttt{number of estimators} from 10 to 100, with 100 being the best value, \texttt{maximum tree depth} from 20 to 70, with 40 being the ideal value, \texttt{minimum number of samples to split} from 10 to 50, with 50 being the optimal value, and \texttt{minimum number of samples to be at a leaf} from 10 to 50, with 13 being the best value. To explore the hyperparameter space, the expected improvement acquisition function is used \citep{zhan2020expected}.\\





\noindent \textbf{Quantile Gradient Boosting}
\label{ss:QGB}

QGB are based on boosting methods that sequentially combine an ensemble of weak learners as a weighted sum of base-learner models to reduce the ensemble error~\citep{friedman2001}. Integral to its probabilistic forecasting capability is the implementation of the pinball loss function, which is similarly employed in QLR to facilitate the generation of probabilistic predictions~\citep{Verbois2018}. The optimization of hyperparameters was performed as QRF, with the inclusion of the \texttt{learning rate} = [0.05, 0.1, 0.5], with the best value of 0.05.

\subsection{Performance Assessment Metrics}

The correctness of probability predictions is assessed through continuous ranked probability score, while prediction robustness is quantified through sharpness and calibration. Finally, prediction interval coverage probability is used to define health-state of the battery.\\

\noindent \textbf{Continuous Ranked Probability Score} (CRPS) can be formally expressed as a quadratic measure of discrepancy between the predicted Cumulative Distribution Function (CDF), $F(\cdot)$, and the observed empirical CDF for a given scalar observation $y$~\citep{zamo2018}:

        \begin{equation}
        \label{eq:crps}
            CRPS(F,y) = \int (F(x) -\mathds{1}(x\geq y_i))^2 dx,
        \end{equation}

\noindent where $\mathds{1}(x\geq y_i)$ is the indicator function, which models the empirical  CDF. 

In order to obtain a single score value from Eq.~(\ref{eq:crps}), a weighted average is computed for each individual observation of the test set~\citep{Gneiting2005}:

    \begin{equation}
    \label{eq:crps_avg}
         CRPS = \frac{1}{N} \sum_{i=1}^{N} CRPS(F_i,y_i)
    \end{equation}

\noindent where $N$ denotes the total number of predictions.\\

\noindent \textbf{Calibration} refers to the statistical consistency between the predictive distributions and the actual observations. It represents a joint property of forecasts and empirical data~\citep{jung2022}. Namely, it is stated that the model is calibrated if~\citep{kuleshov2018}:

    \begin{equation}
        \label{eq:calibration}
        \frac{\sum_{t=1}^{T} \mathds{I}\{y_t \leq F_{t}^{-1}(p)\} }{T} \rightarrow p \text{ for all } p \in [0,1]
    \end{equation}

\noindent In this expression, $T$ refers to the total number of data points, while the indicator function $\mathds{I}\{y_t \leq F_{t}^{-1}(p)\}$ takes a value of 1 when the condition $y_t \leq F_{t}^{-1}(p)$ is true, and 0 otherwise. Given this condition, $y_t$ express the observed outcome at time $t$, and $F_{t}^{-1}(p)$ is the inverse of the CDF for the forecast, evaluated at probability $p$. Therefore, the condition represents the threshold below which a random sample from the distribution would occur with a probability $p$. For example, a well-calibrated model implies that $y_t$ should fall within a 90\% confidence interval around 90\% of the time. \\


\noindent \textbf{Sharpness} means that the confidence intervals should be optimized for minimal width around a singular value. That is, the goal is to reduce the variance, denoted as $var(F_n)$, of the random variable characterized by the cumulative distribution function $F_n$~\citep{kuleshov2018, Tran_2020}:

    \begin{equation}
        \label{eq:sharpness}
        sha = \sqrt{\frac{1}{N} \sum_{n=1}^{N} var(F_n)}
    \end{equation}

\noindent \textbf{Prediction Interval Coverage Probability} (PICP) is a statistical metric that quantifies the reliability of a prediction interval. Namely, it evaluates the proportion of true observations that fall within the estimated prediction interval. This metric provides an assessment of the effectiveness of the prediction interval in capturing the actual variability of the data, thus reflecting the accuracy and validity of the model predictions~\citep{GONZALEZ2021}:
    \begin{equation}
        \label{eq:picp}
        PICP = \frac{1}{N} \sum_{i=1}^{N} c_i
    \end{equation}
    


\noindent where $N$ represents the total number of samples, and $c_i$ equals 1 when the actual observation is within the prediction interval for the specified significance level, and 0 otherwise.

\section{Results}
\label{sec:Results}

The proposed approach is validated in two phases. The first stage focuses on the benchmarking assessment for EOD voltage predictions to obtain the most accurate and robust predictive model. The second stage focuses on battery health-state diagnostics. That is, different batteries are assessed using flight record data to quantify uncertainty and diagnose the health-state.

\subsection{EOD Voltage Prediction}
\label{ss:Results_EODV_prognostics}

Table~\ref{table:Performance} displays the CRPS performance results along with the standard deviation (std) for various probabilistic models tested across a range of real flights (cf. Table~\ref{table:dataset}). 

\begin{table*}[!ht]
    \centering
    \caption{CRPS (std) of different probabilistic models tested across real flights. The best results are in \textbf{bold}.}
        \begin{tabular}{c c c c c c }
             \toprule
             \textbf{Flight (\#)} & \textbf{Sample (\%)} & \textbf{QLR} & \textbf{QRF}& \textbf{QGB}& \textbf{CNN MC dropout} \\
           \midrule
            11    & 14.14 & 0.023 (0.007)   & 0.028 (0.021)  & 0.024 (0.019)   & \textbf{0.022(0.003)}   \\
            30    & 15.92 & 0.037 (0.013)   & 0.037 (0.024)  & 0.039 (0.033)   & \textbf{0.023(0.003)}   \\
            39    & 15.51 & 0.017 (0.005)   & 0.027 (0.023)  & 0.021 (0.027)   & \textbf{0.016(0.006)}   \\
            44    & 3.99 & 0.031 (0.009)   & 0.036 (0.021)  & 0.033 (0.019)   & \textbf{0.026(0.003)}   \\
            69    & 13.83 & 0.029 (0.011)   & 0.034 (0.022)  & 0.035 (0.031)   & \textbf{0.023(0.002)}   \\
            73    & 14.71 & 0.031 (0.009)   & 0.036 (0.024)  & 0.038 (0.034)   & \textbf{0.023(0.003)}   \\
            85    & 13.41 & 0.045 (0.014)   & 0.066 (0.037)  & 0.049 (0.056)   & \textbf{0.034(0.008)}   \\
            88    & 8.49 & 0.027 (0.007)   & 0.032 (0.019)  & 0.035 (0.031)   & \textbf{0.022(0.002)}   \\ 
            \midrule
            \textbf{TOTAL}&100 &0.027 (0.009)& 0.037 (0.023) &0.034 (0.031) &\textbf{0.023 (0.003)}\\
            \bottomrule
        \end{tabular}
        \label{table:Performance}
\end{table*}

Namely, the CRPS is evaluated for each test sample of each flight, \textit{i.e.} the whole flight contains a set of CRPS. This set is evaluated as a Gaussian probability distribution function, characterized by the mean and standard deviation. Accordingly, the lower the standard deviation, the greater the consistency between the predicted probability distribution and the actual observed value. To avoid confusions, it is important to remark that the standard deviation does not represent the spread of the predictive distribution.


It can be observed that the CNN MC dropout outperforms the other probabilistic methods. It has the lowest predictive error, and in addition, it has the lowest standard deviation. This supports the idea that the predictions of this model are more consistent with a higher degree of robustness. 

\begin{figure*}[!ht]
    \centering
    \subfigure[]{\includegraphics[width=0.49\textwidth]{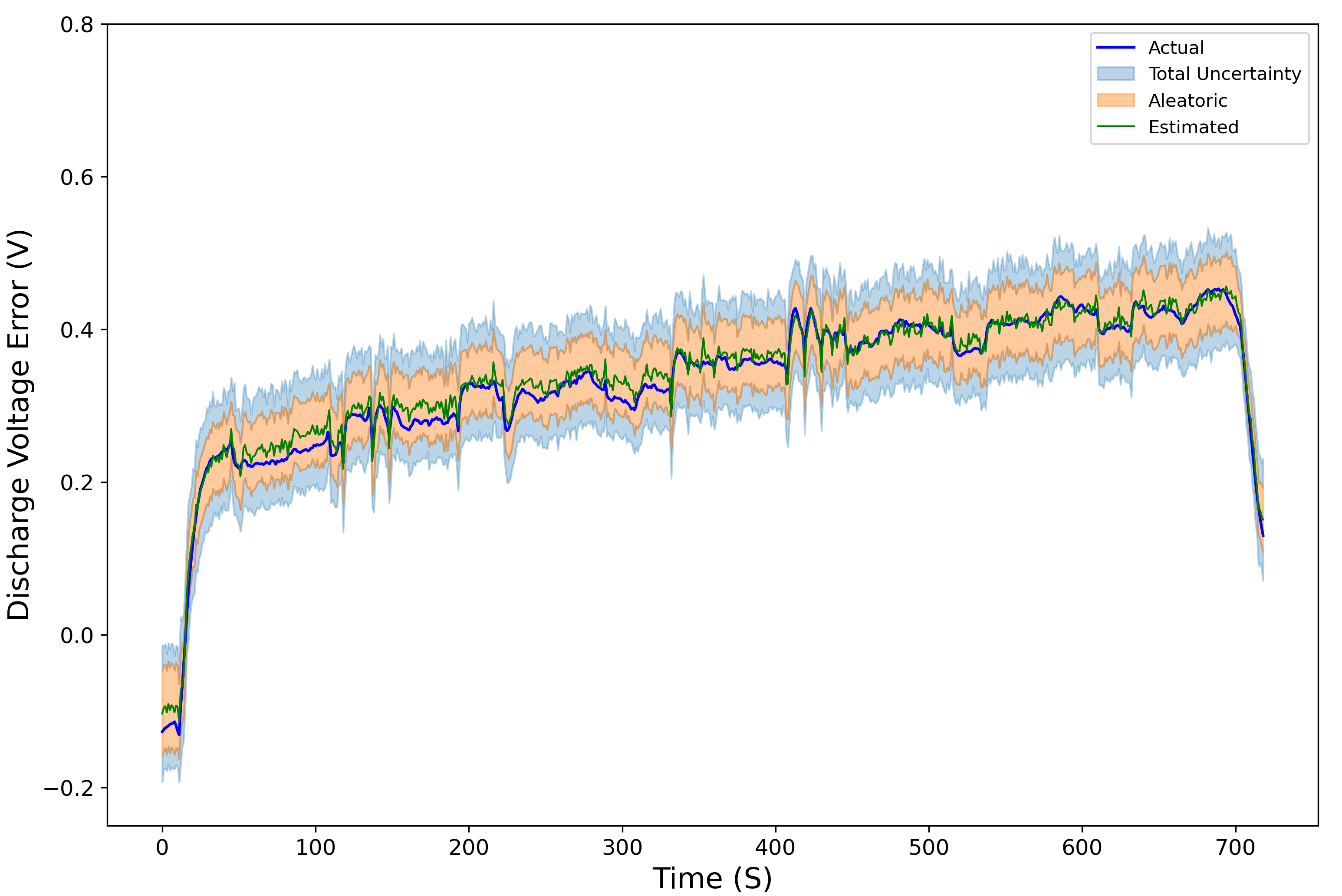}\label{subfig:flight73_cnn}} 
    \subfigure[]{\includegraphics[width=0.49\textwidth]{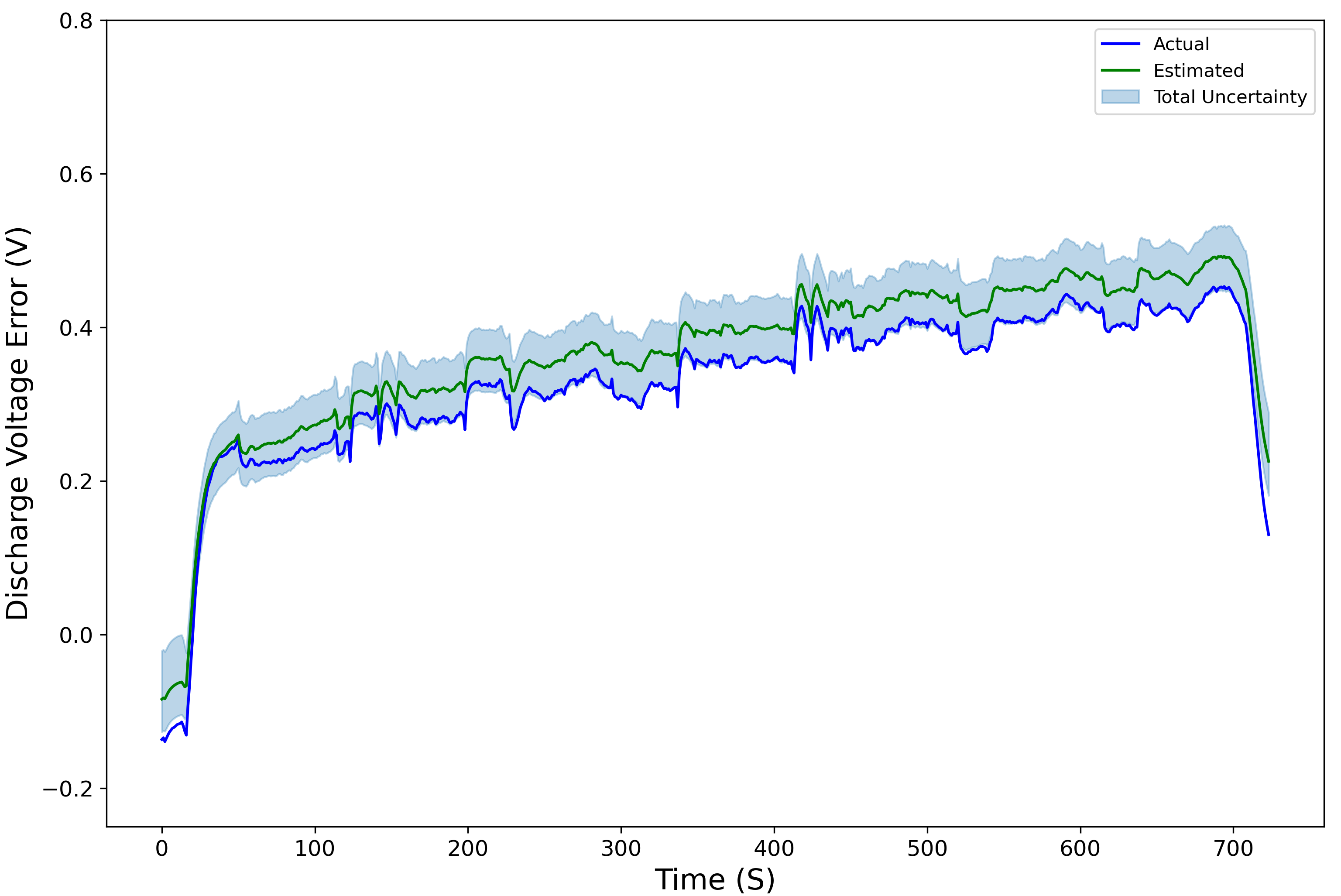}\label{subfig:flight73_qlr}} 
    \subfigure[]{\includegraphics[width=0.49\textwidth]{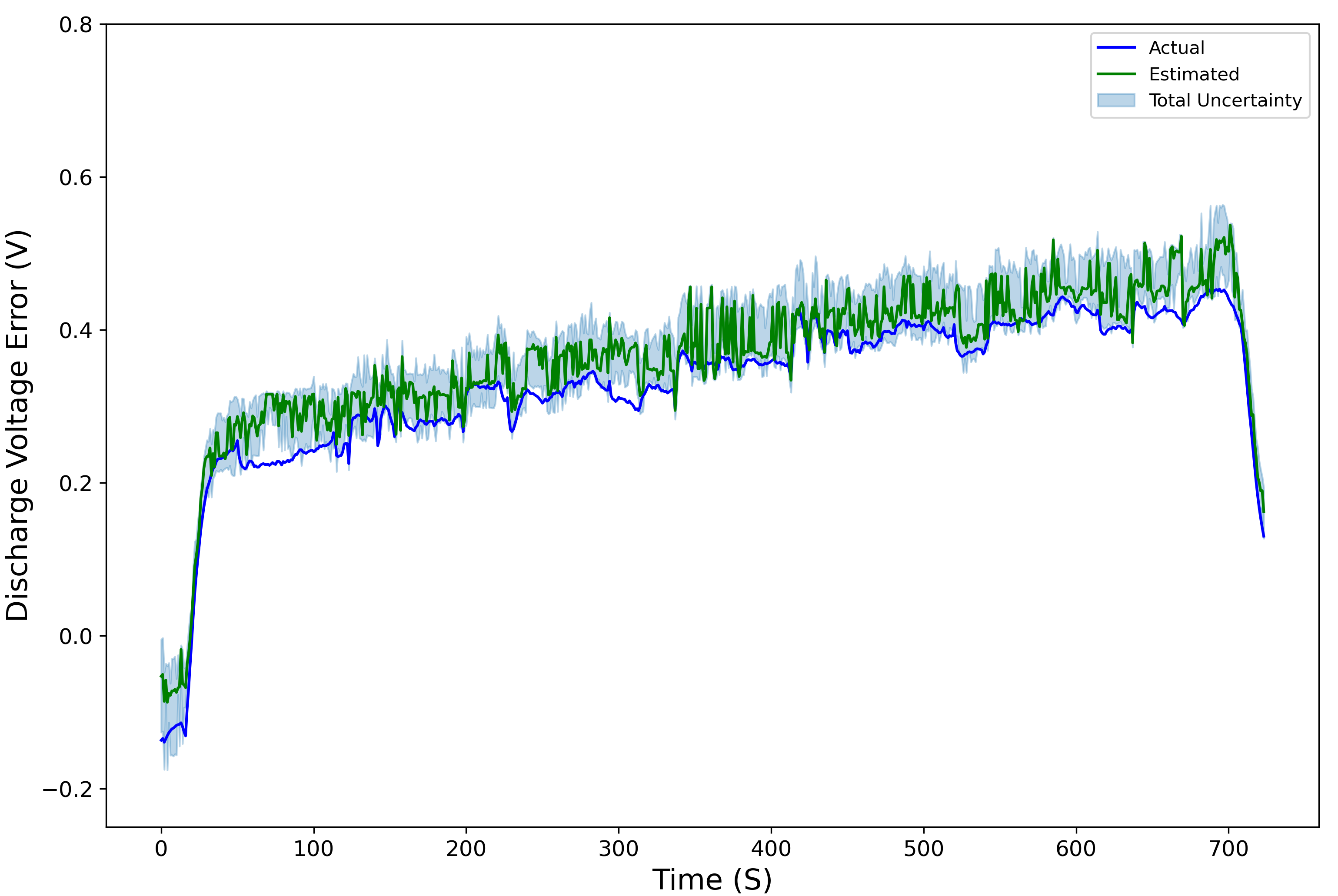} \label{subfig:flight73_qrf}}
    \subfigure[]{\includegraphics[width=0.49\textwidth]{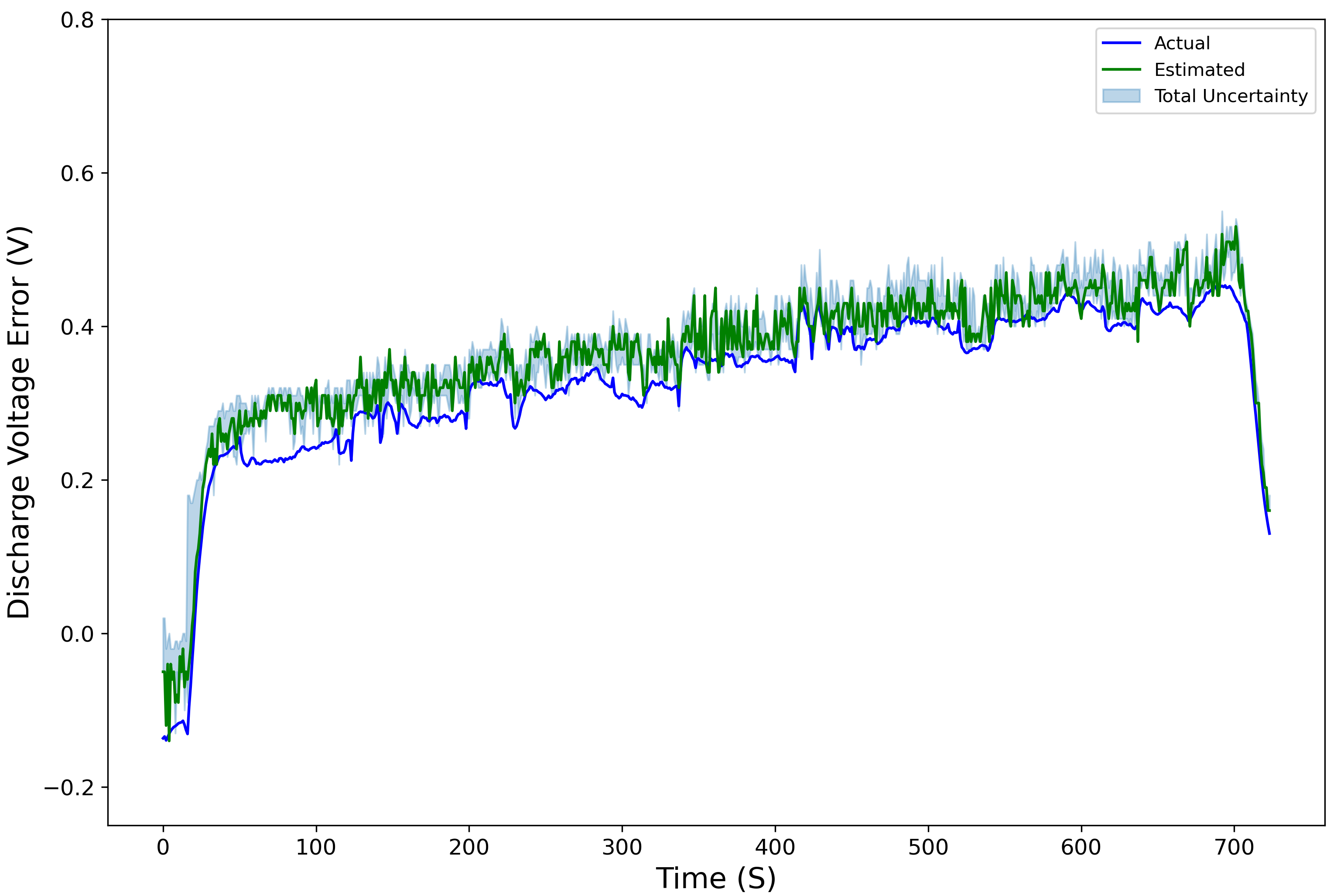}  \label{subfig:flight73_qgb}}
    \caption{Error correction performance across different models for flight \#73 (a) Error correction with CNN Dropout (b) Error correction with QLR (c) Error correction with QRF (d) Error correction with QGB}
    \label{fig:error_correction_results}
\end{figure*}

Figure~\ref{fig:error_correction_results} shows the error prediction example for an individual flight using different data-driven error-correction techniques. It can be observed that (i) all the hybrid prediction models track the error voltage discharge dynamics correctly and (ii) all the results include different sources of uncertainty modeled as probabilistic error estimates.

\begin{figure}[!ht]
    \centering
    \includegraphics[width=0.6\textwidth]{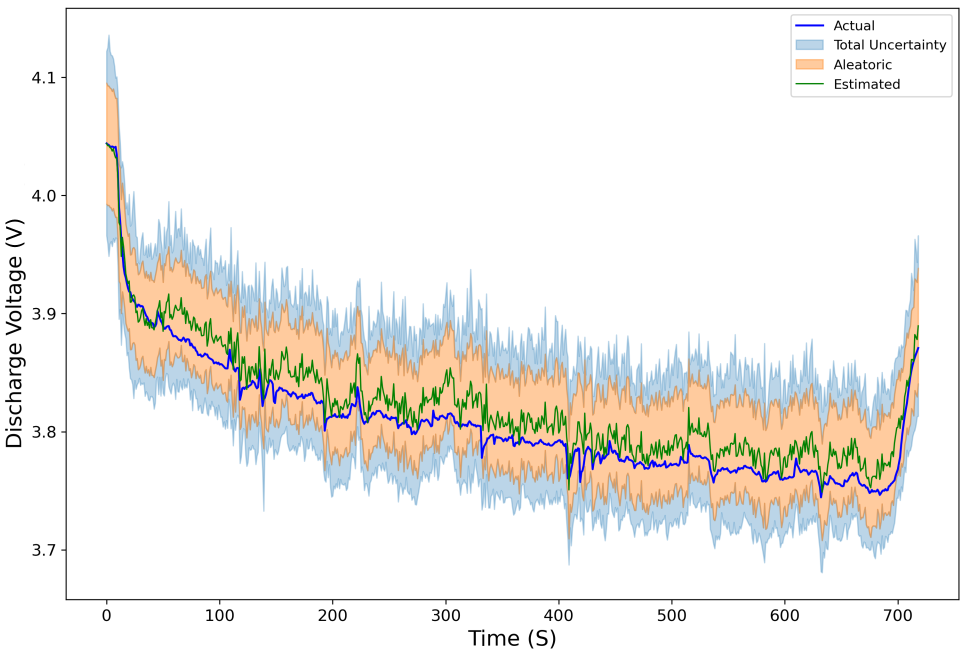}
    \caption{Estimation of EOD Voltage of a battery cell in Flight \#73 using CNN MC dropout.}
    \label{fig:Flight_73_Total_Unc_6S}
\end{figure}

The performance results displayed in Table~\ref{table:Performance} are aligned with flight results in Figure~\ref{fig:error_correction_results}. Namely, CNN MC dropout exhibits higher probabilistic accuracy and reliability when assessing error voltage estimation and the uniformity of predictions, in comparison to the alternative models [cf. Figure~\ref{subfig:flight73_cnn}]. Additionally, the uniformity in the dispersion of the predictive results, as evidenced in Table~\ref{table:Performance}, is shown with a reduced standard deviation of 0.009 volts. On the other hand, Figures~\ref{subfig:flight73_qrf} and ~\ref{subfig:flight73_qgb} present the highest prediction errors, and accordingly, the highest level of variance, in agreement with Table~\ref{table:Performance} (Flight \#73, QRF, QGB). Based on these prediction errors, Figure~\ref{fig:Flight_73_Total_Unc_6S} displays the complete voltage drop of a single-cell battery with the associated uncertainty.

Evaluating the PDF is a crucial aspect for uncertainty quantification. Accordingly, the calibration and the sharpness assessment of PDFs is performed through a python toolbox for predictive uncertainty quantification~\citep{chung2021uncertainty}.

\begin{figure*}[!ht]
    \centering
    \subfigure[]{\includegraphics[width=0.49\textwidth]{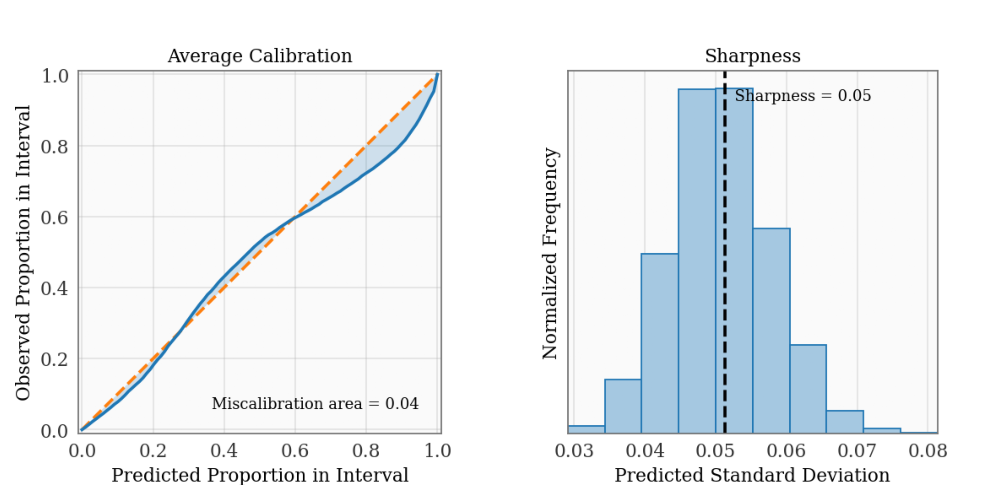}\label{subfig:cal_sharp_cnn}} 
    \subfigure[]{\includegraphics[width=0.49\textwidth]{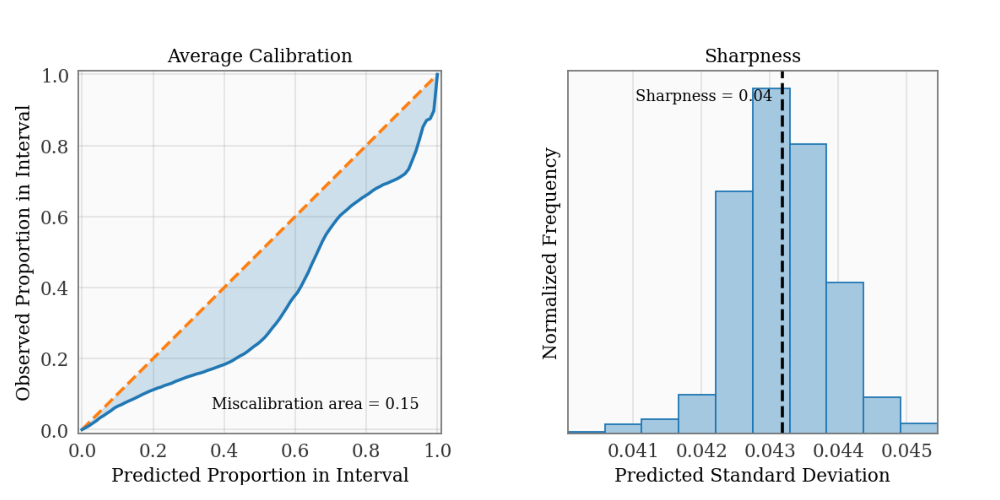}\label{subfig:cal_sharp_qlr}} 
    \subfigure[]{\includegraphics[width=0.49\textwidth]{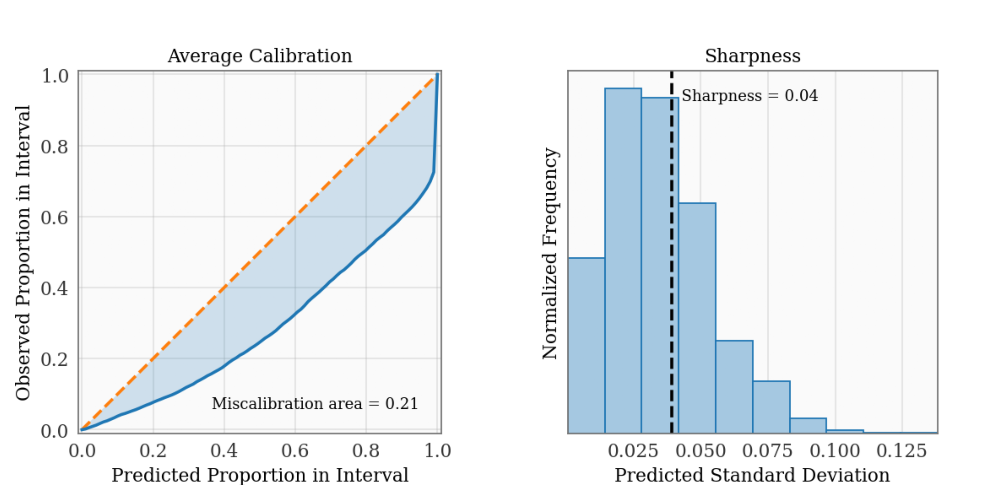}\label{subfig:cal_sharp_qrf}}
    \subfigure[]{\includegraphics[width=0.49\textwidth]{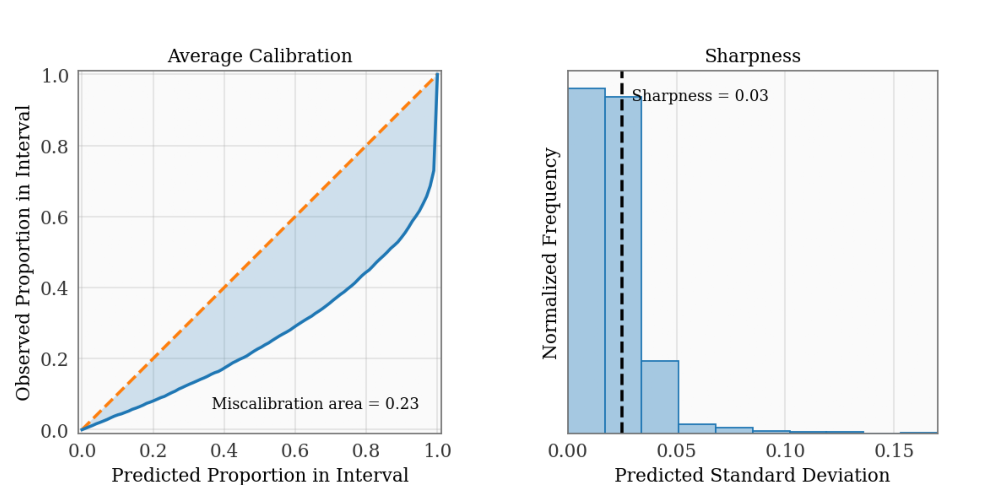}\label{subfig:cal_sharp_qgb}}
    \caption{Calibration and sharpness metrics for all UQ methods used in this study. (a)  CNN MC dropout model (b) QLR model (c) QRF model (d) QGB model.}
    \label{fig:model_cal_sharp}
\end{figure*}

Figure~\ref{fig:model_cal_sharp} shows the calibration plot of the developed models, where the X-axis represents the predicted probabilities generated by the model and the Y-axis represents the observed outcomes. A well-calibrated model should have points on the diagonal line, indicating that the predicted probabilities match the observed datapoints. Therefore, the model calibration in solid blue lines can be determined by assessing the distance of the calibration curves to the diagonal line. This is quantified by calculating the area between the calibration curve and the diagonal line, represented in blue shaded area, \textit{i.e.} miscalibration area. This metric is characterized by smaller values, which indicate improved calibration~\citep{Tran_2020}. Based on the miscalibration area, the CNN MC dropout model [Figure~\ref{subfig:cal_sharp_cnn}] has the lowest miscalibration area of 0.04 which suggest a well calibrated model. This model demonstrates excellent calibration accuracy. In contrast, both the QRF and QGB models exhibit higher miscalibration areas of 0.21 and 0.23, respectively, indicating miscalibration. Finally, the QLR model presents an acceptable calibration, exhibiting a miscalibration area of 0.15 in contrast to other quantile methods.

Simultaneously, sharpness is also evaluated, which quantifies the concentration of distributional predictions~\citep{Tran_2020}. In this case, as the predictions parameterize a Gaussian distribution, the variance of the predicted distribution is commonly considered as a metric for sharpness. Generally, there is a trade-off between calibration and the sharpness of predictive distribution ~\citep{Gneiting2007}. Figure~\ref{fig:model_cal_sharp} shows all predictive uncertainty distributions of the probabilistic models, where the sharpness values are indicated by vertical lines.

The results show that QGB exhibits the lowest degree of sharpness [Figure~\ref{subfig:cal_sharp_qgb}] implying a low level of spread. However, it is worth noting that this model corresponds to the poorest model calibration, as previously observed. Consequently, it becomes evident that achieving a trade-off between calibration and sharpness is imperative. In this context, QLR emerges as a model that effectively maintains the trade-off between these two metrics, achieving a sharpness of 0.04V [Figure~\ref{subfig:cal_sharp_qlr}].

On the other hand, despite the CNN MC dropout displays the highest sharpness, the difference respect other models is minimal [Figure~\ref{subfig:cal_sharp_cnn}]. The sharpness presented for CNN MC dropout stands at 0.05V, representing a marginal increase of 0.01V with respect to QLR. In addition, this difference suggests that the sharpness observed in the CNN MC dropout model does not come at a significant calibration cost.

Obtained results, confirm the superior CRPS performance, calibration, and sharpness of  the proposed probabilistic hybrid model validate   the contribution and illustrate the EOD voltage prediction effectiveness  of the proposed methodology.



\subsection{Health-State Diagnostics}
\label{ss:Results_Health-StateDiagnostics}
Table~\ref{table:PICP_Performance} shows the health-state estimations from eight individual flight instances of different drones equipped with different batteries of same capacity attributes. In particular, each flight of this testing set is equipped with different Li-Po batteries with same capacity attributes. The qualitative assessment of each battery degradation was determined by expert personnel categorizing the batteries into good, medium, and bad health according to their usage history and operation conditions. However, this classification lacks precise quantification features. To quantify degradation using total predictive uncertainty, a numerical scale is utilized (cf. Figure~\ref{fig:health_state}), \textit{i.e.}, 0 indicates batteries in degraded health-state and 1 denotes optimal health-state batteries.

\begin{table}[!ht]
    \centering
    \caption{Health-state estimation of different models for different real flights equipped with batteries in different health conditions.}
    \setlength{\tabcolsep}{2pt}
        \begin{tabular}{c c c c c }
             \toprule
             \textbf{Flight (\#)}  & \textbf{QLR} & \textbf{QRF}& \textbf{QGB}& \textbf{CNN MC dropout} \\
           \midrule
            11 & \gradient{0.956}{nine}{ten} & \gradient{0.806}{eight}{nine} & \gradient{0.601}{six}{seven} & \gradient{0.998}{nine}{ten}\\
            30 & \gradient{0.602}{six}{seven} & \gradient{0.471}{four}{five} & \gradient{0.191}{one}{two} & \gradient{0.993}{nine}{ten}\\
            39 & \gradient{1.0}{nine}{ten} & \gradient{0.841}{eight}{nine} & \gradient{0.605}{six}{seven} & \gradient{0.997}{nine}{ten}\\
            44 & \gradient{0.938}{nine}{ten} & \gradient{0.903}{nine}{ten} & \gradient{0.541}{five}{six} & \gradient{1.0}{nine}{ten}\\
            69 & \gradient{0.834}{eight}{nine} & \gradient{0.659}{six}{seven} & \gradient{0.306}{three}{four} & \gradient{1.0}{nine}{ten}\\
            73 & \gradient{0.763}{seven}{eight} & \gradient{0.506}{five}{six} & \gradient{0.157}{one}{two} & \gradient{1.0}{nine}{ten}\\
            85 & \gradient{0.283}{two}{three} & \gradient{0.184}{one}{two} & \gradient{0.068}{one}{two} & \gradient{0.948}{nine}{ten}\\
            88 & \gradient{0.961}{nine}{ten} & \gradient{0.727}{seven}{eight} & \gradient{0.241}{two}{three} & \gradient{1.0}{nine}{ten}\\
            \bottomrule
        \end{tabular}
        \label{table:PICP_Performance}
\end{table}	

Conducting a detailed analysis of the results in Table~\ref{table:Performance}, it becomes evident that the UQ modeling using QGB methods falls short in accurately identifying the true health-state of batteries used in each flight. However, the QRF method outperforms in this regard, showing notably improved results in identifying the batteries associated with flights \#11, \#39, \#44, and \#88.

On the other hand, QLR method, demonstrates a robust health-state index modeling for most batteries, with the exception of flights \#30 and \#85. In these instances, a significant portion of observations falls outside the uncertainty interval, indicating a suboptimal battery health condition.

Finally, the CNN MC dropout exhibits the most effective uncertainty modelling, accurately estimating all health-state estimation values for the different batteries as confirmed by expert personnel. The combination of aleatoric and epistemic uncertainty provides precise information, enabling the probabilistic CNN to generate a precise uncertainty-aware health indicator. The proposed  uncertainty quantification approach supports our final contribution, indicating that addressing both aleatoric and epistemic uncertainty improves the final decision-making process.

\section{Discussion}
\label{sec:Discussion}

A comprehensive hybrid probabilistic battery health management approach for robust drone inspections is presented in this research. However, before drawing definite conclusions, there are some key areas of the developed research that are worth discussing.



\subsection{Robustness analysis}
\label{ss:disc_Robustness}

\begin{figure*}[!ht]
    \centering
    \includegraphics[width=0.62\textwidth]{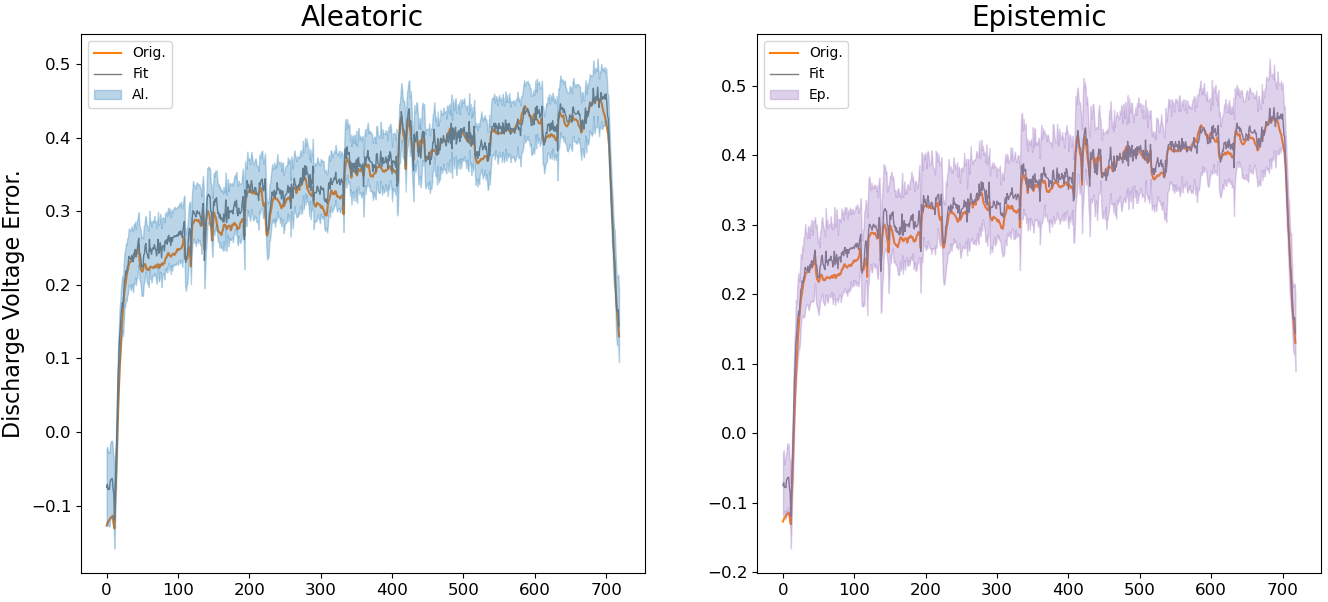}
    \caption{Estimation of EOD Voltage Correction in Flight \#73 using CNN MC dropout for the Discrimination of Aleatoric and Epistemic Uncertainty.}
    \label{fig:Flight_73_Uncertainty}
\end{figure*}

The proposed MC dropout based CNN model demonstrates a superior performance in terms of accuracy and uncertainty quantification (cf. Section \ref{ss:Results_EODV_prognostics}). Importantly, the proposed approach can quantify aleatoric and epistemic uncertainty, as shown in Figure~\ref{fig:Flight_73_Uncertainty}, where the blue area denotes the uncertainty inherent in the data, often arising from sensor noise (aleatoric), and the purple area represents the uncertainty generated by the model itself (epistemic). 

A more detailed analysis reveals that the epistemic uncertainty exceeds the aleatoric uncertainty. This suggests that epistemic uncertainty reduction methods, e.g. optimizing the model calibration or augmenting the training dataset, may be needed to mitigate the main source of uncertainty in this case. The explicit differentiation between different sources of uncertainty is a key outcome of the proposed approach. The proposed UQ approach provides valuable insights, which contribute towards a more robust and accurate understanding of uncertainty compared with other probabilistic methods (cf. Section \ref{sec:Results}).

\subsection{Sensitivity analysis of dropout rate}
\label{ss:disc_sensisitivity}

In order to evaluate the influence of the dropout rate on the obtained results, this section discusses the influence of the dropout rate on different metrics and the impact on epistemic uncertainty. Taking the flight \#73 as a reference, firstly, the influence of the variations in the dropout rate on the calibration, sharpness and CRPS metrics have been evaluated. Table~\ref{table:DropoutRate_sens} displays the obtained results.

\begin{table}[!htb]
    \centering
    \caption{Sensitivity analysis of dropout rate for the flight \#73.}
        \begin{tabular}{c c c c }
           \toprule
             \textbf{Dropout Rate} & \textbf{Calibration} & \textbf{Sharpness} & \textbf{CRPS} \\
           \midrule
            0.01    & 0.06 & 0.05   & 0.0178   \\
            0.05    & 0.09 & 0.09   & 0.0267   \\
            0.1     & 0.24 & 0.18   & 0.0482   \\
            0.15    & 0.28 & 0.21   & 0.0503   \\
            0.2     & 0.29 & 0.23   & 0.0561   \\
            \bottomrule
        \end{tabular}
        \label{table:DropoutRate_sens}
\end{table}	

It can be observed that the dropout rate has a significant impact on calibration, sharpness, and CRPS metrics. Namely, incremental increases in the dropout rate from 0.01 to 0.2 lead to notable improvements in these metrics, with calibration values increasing from 0.06 to 0.29, sharpness from 0.05 to 0.23, and CRPS from 0.0178 to 0.0561. 



\begin{figure}[!ht]
    \centering
    \includegraphics[width=0.65\linewidth]{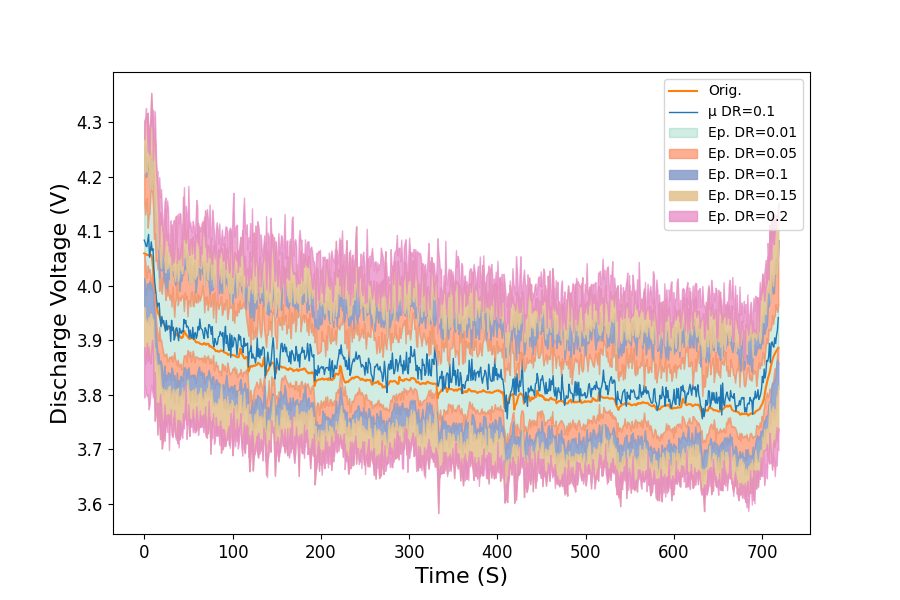}
    \caption{Epistemic uncertainty variation for the flight \#73 due to incremental increases in dropout rate.}
    \label{fig:DropoutRate_epistemic}
\end{figure}

Subsequently, the influence of the dropout rate on the epistemic uncertainty has been evaluated as shown in Figure~\ref{fig:DropoutRate_epistemic}. It can be observed that, as the dropout rate increases, epistemic uncertainty also rises.

Consequently, the selected dropout rate value in the experiments ($\rho$=0.1) is a balanced trade-off decision between overfitting and high model uncertainty.


\subsection{Model complexity and accuracy}
\label{ss:disc_tradeoff}

The proposed approach has shown superior performance in terms of accuracy and uncertainty quantification. However, the model complexity should also be considered for a fair evaluation. 

Namely, the trade-off between model complexity and predictive accuracy was observed when QLR was compared with MC dropout based CNN. QLR, is a simpler model with fewer parameters (12 parameters), exhibited lower computational costs for training and inference (0.00045 ms/step) for a standard desktop Core i5 10th Gen. However, this simplicity came at the cost of predictive accuracy, as QLR underperformed compared to the CNN model (cf. Table~\ref{table:Performance}). 

On the other hand, MC dropout CNN, characterized by their complex architectures and large number of tunable parameters (4912 parameters), delivered superior performance in terms of accuracy. Nevertheless, the computational cost associated with CNNs was significantly higher (0.065 ms/step) due to several iterations to predict the distribution, requiring more processing power and memory. 


\subsection{Health-state estimation vs uncertainty}
\label{ss:disc_uncertainty}

The proposed health-state estimation is focused on a probabilistic distance-metric. The approach infers battery health-state based on probabilistic prediction uncertainty with respect to ground truth.
 
There are other health-state estimation specific methods that elicit a health index which satisfies prognostics-specific properties \citep{Zhen_23}. However, the main goal of this research is to develop a robust battery health management approach based on accurate uncertainty-aware discharge voltage predictions. To demonstrate the use and validity of the modelled uncertainty, an application has been developed focused on health-state estimation.

\subsection{Limitations}
\label{ss:disc_limitations}

This research, however, is subject to several limitations. On the one hand,the computational demand for real-time operations in drones may be excessive due to the  demanding computational power of CNNs and MC dropout. On the other hand, the operational conditions covered by the employed dataset may be limited, and they may not include a complete set of diverse operation conditions, such as irregular or extreme operating conditions. Depending on the operation context, this  may lead to an underestimation of the predictive accuracy of the model.

\section{Conclusion}
\label{sec:Conclusion}

This research introduced a drone battery health assessment approach that combines physics-based models with probabilistic data-driven error prediction techniques. Both methods have been integrated into a hybrid error correction configuration. The proposed methodology demonstrated robust performance in the accurate prediction of battery discharge voltage and effectively quantified uncertainties originated from data and the model. 

The performance of the hybrid probabilistic methodology was empirically evaluated on a dataset comprising EOD voltage under varying load conditions. The dataset was generated using real inspection drones on different flight missions, focused on offshore wind turbine inspections. Probabilistic CNN demonstrates a 14.8\% enhancement in probabilistic accuracy over QLR in battery discharge voltage predictions, with 37.8\% and 32.3\% improvements for QRF and QGB, respectively. In addition, aleatoric and epistemic uncertainties provide robust estimations to enhance the diagnosis of battery health-states.

In the conducted experiments, the proposed hybrid prediction framework exhibited an accurate performance, even in the presence of uncertainties associated with the data and the model. In addition, the study demonstrated that the hybrid error correction methodology can achieve a high prediction performance when applied to Li-Po batteries, despite of a Li-Ion physics-based model.

Future research will focus on addressing computational power and alternative hybrid configurations through several strategies. Computational efficiency may be improved using techniques like quantization~\citep{cheng2018}, tensor factorization~\citep{kim2016}, and low-rank approximation~\citep{tai2016}. Benchmarking different methods and integrating the features of different uncertainty quantification (UQ) methods will help developing a more robust tool for drone battery health assessment. In this direction, PINNs variants will be explored as alternative hybrid configurations. By incorporating physical laws into the learning process, PINNs will be benchmarked against the proposed approach to assess their efficacy in enhancing the robustness and reliability of drone battery health predictions. Alternatively, the dataset may be extended including data from batteries under a diverse set of operational conditions, including features beyond battery capacity, such as controlled charging and discharging cycles. Additionally, future drones will be equipped with built-in sensors to monitor environmental conditions, which will be incorporated into the battery model, enabling more accurate predictions. Future work will also explore drone path planning considering interactions with the operational environments and elicitation of optimal operation policies. Reinforcement Learning strategies may offer promising directions for adapting drone operations based on battery health and environmental factors~\citep{GAO2024}.

Moreover, a possible avenue to improve the prediction accuracy may be focused on the use of calibration methods, such as isotonic regression~\citep{kuleshov2018}, and comparing with other models like Bayesian Neural Networks to integrate the best features of various UQ methods~\citep{alcibar2024}.

\section*{Acknowledgements}

This publication is part of the research projects KK-2022-00106, KK-2023-00041, IT1451-22 and IT1676-22 funded by the Basque Government. J. I. Aizpurua is funded by Juan de la Cierva Incorporacion Fellowship, Spanish State Research Agency (grant No. IJC2019-039183-I). The authors would like to thank Matteo Corbetta at KBR, Inc. NASA Ames Research Center for his comments and useful discussions.



 \bibliographystyle{elsarticle-harv}\biboptions{authoryear,round}

\end{document}